\documentclass[journal]{IEEEtran}
\usepackage{amsmath,amsfonts,amsthm,amssymb}
\usepackage{graphicx}
\usepackage{textcomp}
\usepackage{mdwtab}
\usepackage{eqparbox}
\usepackage{bbm}
\usepackage{array}
\usepackage{textcomp}
\usepackage[caption=false,font=footnotesize]{subfig}%
\usepackage{url}
\usepackage{verbatim}
\usepackage{cite}
\usepackage{multirow}
\usepackage{indentfirst}
\usepackage{booktabs}
\usepackage{breqn}
\usepackage{accents}
\usepackage{bm}
\usepackage{makecell}
\usepackage{nomencl}
\usepackage{mathtools}
\usepackage{enumerate}
\usepackage[caption=false,font=footnotesize]{subfig}
\usepackage{dsfont}
\usepackage[ruled]{algorithm2e}
\usepackage{algpseudocode}
\allowdisplaybreaks 
\usepackage[subtle,tracking=normal]{savetrees}

\def\BibTeX{{\rm B\kern-.05em{\sc i\kern-.025em b}\kern-.08em
    T\kern-.1667em\lower.7ex\hbox{E}\kern-.125emX}}
\usepackage{balance}

\theoremstyle{definition}

\newtheorem{theorem}{Theorem}
\newtheorem{proposition}{Proposition}

\usepackage{url}
\usepackage{hyperref} 
\usepackage{xcolor}

\hypersetup{
    colorlinks=true, 
    linkcolor=blue, 
    filecolor=black, 
    urlcolor=blue, 
    citecolor=blue, 
    linkbordercolor={1 1 1}, 
    pdfborder={0 0 0} 
}

\begin{document}

\title{Real-time Coordination of Cascaded Hydropower \\ under Decision-Dependent Uncertainty}

\author{Eliza Cohn, \IEEEmembership{Student Member, IEEE}, Ning Qi, \IEEEmembership{Member, IEEE}, Upmanu Lall, and Bolun Xu, \IEEEmembership{Member, IEEE}

\thanks{This work was supported by the National Science Foundation Graduate Research Fellowship under Grant No. DGE-2036197 and by
the Department of Energy under Grant No. DE-EE0011385. \textit{(Corresponding author: Ning Qi)}

Eliza Cohn, Ning Qi, and Bolun Xu are with the Department of Earth and Environmental Engineering, Columbia University, New York, NY 10027 USA (e-mail: \{ec3766,nq2176,bx2177\}@columbia.edu). 

Upmanu Lall is with the School of Complex Adaptive Systems, Arizona State University, Tempe, AZ 85287 USA (e-mail: ulall@asu.edu).}}

\markboth{IEEE Transactions on Sustainable Energy, VOL. XX, NO. XX, XXXX}
{How to Use the IEEEtran \LaTeX \ Templates}
\maketitle


\begin{abstract}
This study proposes a real-time control framework for cascaded hydropower systems that incorporates decision-dependent uncertainty (DDU) to capture the coupling of streamflow uncertainties across reservoirs. The framework jointly models exogenous forecast errors and endogenous uncertainty propagation, characterizing the dependence between upstream releases and downstream inflow variability through a heteroskedastic variance model conditioned on past forecast errors and releases. We formulate a rolling-horizon joint chance-constrained optimization under DDU to ensure reliable cascade hydropower dispatch, with storage-reference tracking to address the future water value and a McCormick relaxation of the nonconvex head. We develop a tractable supporting hyperplane algorithm that decouples uncertainty from decisions across time steps and directly enforces the joint chance constraint under DDU at the prescribed risk level. We establish the convergence of the proposed method and characterize its adaptive risk allocation behavior. Case studies on the Lower Columbia River demonstrate that incorporating DDU reduces the integrated violation index by up to 34.2\% and increases total generation by up to 12.8~GWh over a year relative to decision-independent uncertainty (DIU). Sensitivity and scalability analyses across Monte Carlo inflow scenarios further demonstrate the value of the framework for reliable real-time cascaded hydropower operations.

\end{abstract}


\section{Introduction}\label{sec:introduction}

Hydropower accounts for 27\% of U.S. renewable electricity generation and provides essential water-management services, including flood control, irrigation, and water supply~\cite{Rocio-Uria-Martinez2023-fw}, making it a critical component of the energy--water nexus and a flexible low-carbon resource in the evolving power system~\cite{Kan2024-ieecsc}. Cascaded hydropower represents a major class of hydropower resources, where multiple units along a river are hydraulically coupled through streamflow propagation and reservoir storage dynamics. Hydropower operations have traditionally relied on long-term planning over monthly or seasonal horizons. However, growing participation in real-time electricity and ancillary service markets~\cite{su2025real,shi2026coordinated} demands dispatch policies at hourly timescales that ensure reliable generation while meeting multi-purpose water-management constraints.

Real-time cascaded hydropower dispatch faces the challenge of coordinating upstream and downstream releases under exogenous and endogenous uncertainties. Exogenous uncertainty arises from weather variability and seasonal hydrologic conditions, which affect natural inflows independently of system operations. Endogenous uncertainty is fundamentally different as upstream release decisions reshape the downstream inflow distribution: travel time and channel storage shift the mean inflow, while streamflow forecast errors grow with flow magnitude~\cite{Schoups2010-wrr} and travel times vary with flow~\cite{Ge2018-km}. Capturing these uncertainties requires modeling the spatial-temporal coupling of streamflows across the cascade, a task that has been approached differently by the hydrology and power system communities. Hydrological studies model streamflow uncertainties through stochastic processes such as nonparametric Markovian simulations~\cite{sharma1997streamflow}, Bayesian frameworks~\cite{Liu2021-cg}, ensemble forecasts~\cite{Faber2001-au}, and integrated forecasting–optimization systems that capture spatial–temporal correlations~\cite{8894506}. However, the computational cost of detailed hydrological models can limit their direct integration into short-term, large-scale dispatch~\cite{Ding2021-ny}.

To address the tractability challenge, hydropower dispatch adopts either deterministic forecasts or decision-independent uncertainty (DIU) models~\cite{Kong2020-la}. Water propagation between cascaded reservoirs has been represented using fixed travel times~\cite{Tong2013-ne}, deterministic river routing~\cite{Diniz2015-rc}, flow-dependent delays~\cite{Ge2018-km}, and Muskingum routing~\cite{Ji2021-be}. With these models, cascade hydropower dispatch is normally formulated as two-stage optimizations, including robust~\cite{Ju2023-ht}, stochastic~\cite{Santosuosso2025-qd}, chance-constrained~\cite{Zhang2022-mu}, and distributionally robust optimization~\cite{Liu2025-ju}. Two-stage methods typically focus on day-ahead scheduling. However, when all intraday decisions are grouped into a single recourse stage, the decision for an early interval may depend on inflows not yet observed, so the recourse decisions need to be updated in real time as new information arrives. Multi-stage methods such as stochastic dynamic programming (SDP) and stochastic dual dynamic programming (SDDP) preserve non-anticipativity through sequential decision making. SDP discretizes the reservoir states and thus suffers from the curse of dimensionality as the number of reservoirs grows~\cite{wu2018stochastic}. SDDP avoids this discretization by approximating the future cost function with sampled cutting planes~\cite{helseth2022hydropower}, but requires convex stage problems, so nonconvex hydropower models need integer extensions~\cite{8289405}. Both compute their policies from uncertainty representations estimated offline, limiting their direct application to real-time dispatch where forecasts must be continuously updated~\cite{Jin2025-od}.

These limitations motivate online optimization methods that leverage continuously updated uncertainty information from forecasts or real-time observations. Model predictive control (MPC) is widely used for real-time hydropower dispatch, repeatedly solving a receding-horizon optimization problem with updated forecasts~\cite{Hamann2017-al,su2025real,Cai2025-pcmp}. However, the performance of MPC depends heavily on forecast quality. Other online approaches include reinforcement learning, which learns dispatch policies from data to enable fast online inference~\cite{Zhang2025-tg,Yang2025-ece,Lu2025-ece}, but requires extensive offline training and offers no explicit guarantee that reservoir limits are met at a prescribed reliability level. Online convex optimization updates decisions sequentially with provable regret bounds~\cite{qi2026online}, but these guarantees rely on convex models. All of these approaches, offline or online, do not explicitly capture this release-dependent inflow uncertainty. Neglecting this endogenous coupling can underestimate downstream forecast risk in real-time cascaded operation.

Decision-dependent uncertainty (DDU) models provide a tractable framework that characterizes how decisions shape uncertainty distributions. Recent work has applied DDU to power system planning and operations through robust~\cite{Zhang2022-yy,chen2022robust}, chance-constrained~\cite{qi2023chance, Zhao2026-qk}, and distributionally robust~\cite{Moreira2024-dr,pianco2025decision} formulations. Among these formulations, chance-constrained optimization suits reservoir dispatch: it meets reservoir limits with a prescribed probability, using a margin set by the release-dependent forecast distribution. Because cascaded reservoirs are hydraulically coupled, their limits should hold jointly, which calls for a joint chance constraint. However, three gaps remain. First, existing DDU formulations typically employ hand-crafted dependency structures or representations learned offline from historical data, which map decisions to a static uncertainty description, whereas in a cascade the forecast variance evolves hour by hour with the latest forecast error and upstream release. Second, most DDU studies are primarily developed for long-term planning or day-ahead dispatch~\cite{pianco2025decision, Zhang2022-yy, qi2023chance, chen2022robust}, leaving real-time dispatch with updated DDU underexplored. Third, the chance-constrained DDU dispatch in~\cite{qi2023chance} handles only individual constraints, while supporting-hyperplane and cutting-plane methods for joint chance-constrained optimization in reservoir management~\cite{van-Ackooij2013-vt} and hydrothermal unit commitment~\cite{van-Ackooij2018-ar} assume DIU.


Beyond uncertainty, real-time dispatch must stay consistent with longer-term plans and handle the model nonconvexity. Long-term planning balances hydropower generation with future water availability and competing water-management objectives~\cite{Giuliani2021-ik,Cohn2025-bg,de-Queiroz2016-il} and values carryover storage~\cite{Chen2025-qm}, while hydrothermal scheduling links this long-term water valuation with detailed day-ahead scheduling and hourly pricing~\cite{Fakhar2021-ui,Santos2020-yr,Diniz2008-fd}. Participation in day-ahead and contract markets further adds opportunity costs to real-time releases~\cite{Hankimaa2025-af}. In real-time dispatch, this future value is typically carried by storage plans or by water values from SDP over Markov-chain models~\cite{Giuliani2021-ik,de-Queiroz2016-il}. Storage plans either from historical reference or day-ahead plan provide a tractable real-time proxy for the future value of water but ignore real-time hydrologic conditions updates, whereas water values inherit the dimensionality limits of SDP and rely on transition probabilities estimated offline. The hydropower generation function is also nonconvex, as it is the product of release and a storage-dependent head. Prior works usually handle this with piecewise linear formulations~\cite{Kong2020-la,Diniz2008-fd}, which are costly to re-solve every hour together with a joint chance constraint.

Table~\ref{tab:literature-comparison} compares the related work and summarizes the research gaps above. Motivated by this, we propose a real-time coordination framework for cascaded hydropower that embeds an hourly updated DDU model of downstream inflow into a joint chance-constrained MPC, with storage-reference tracking to address the future water value and a McCormick relaxation for the nonconvex model. The contributions of this work are:
\begin{enumerate}
    \item \textit{Modeling:} We propose a decision-dependent uncertainty model for cascaded hydropower to explicitly capture how upstream release decisions reshape downstream inflow uncertainty, with both the mean and the variance conditioned on upstream releases and updated every hour.
    \item \textit{Methodology:} We formulate the real-time cascaded hydropower dispatch as a rolling-horizon joint chance-constrained optimization under DDU, with storage-reference tracking to address the future water value and a McCormick relaxation of the nonconvex head. We develop a Sequential Supporting Hyperplane (SSH) algorithm with proved convergence performance that decouples uncertainty from decisions across time steps and directly enforces the joint chance constraint under DDU at the prescribed risk level.
    \item \textit{Numerical study:} We validate the framework with case studies based on the Lower Columbia River, showing that incorporating DDU reduces the constraint violations and increases total generation annually relative to DIU variant and state-of-the-art benchmarks. We test the robustness of the proposed algorithm across Monte Carlo inflow scenarios and its scalability by solving a twelve-unit cascade in under two seconds per horizon.
\end{enumerate}

\begin{table*}[!t]
\centering
\caption{Comparison With Related Work}
\label{tab:literature-comparison}
\footnotesize
\setlength{\tabcolsep}{4pt}
\renewcommand{\arraystretch}{1.0}
\newcommand{\stkc}[1]{\begin{tabular}[t]{@{}c@{}}#1\end{tabular}}
\newcommand{\how}[1]{\\[-1pt]{\scriptsize #1}}
\newcommand{\cell}[2]{\parbox[t]{#1\textwidth}{\raggedright\hyphenpenalty=10000 #2\strut}}
\newcommand{\refs}[1]{\cell{0.20}{\hskip1sp\cite{#1}}}
\begin{tabular}{@{}llccccl@{}}
\hline
\textbf{References} &
\textbf{Uncertainty model} &
\stkc{\textbf{Real-time}\\\textbf{update}} &
\stkc{\textbf{Joint chance}\\\textbf{constraint}} &
\stkc{\textbf{Long--short-}\\\textbf{coordination}} &
\stkc{\textbf{Nonconvex}\\\textbf{head}} &
\textbf{Method} \\
\hline
\refs{Kong2020-la,Tong2013-ne,Diniz2015-rc,Ge2018-km,Ji2021-be,Diniz2008-fd} &
\cell{0.17}{Deterministic} &
$\times$ & $\times$ & $\circ$ & \checkmark &
\cell{0.21}{Mixed-integer linear program} \\
\refs{Giuliani2021-ik,Cohn2025-bg,de-Queiroz2016-il,Chen2025-qm,Fakhar2021-ui,Santos2020-yr,Hankimaa2025-af} &
\cell{0.17}{DIU} &
$\times$ & $\times$ & \checkmark & $\circ$ &
\cell{0.21}{SDP, SDDP} \\
\refs{Zhang2022-mu,Liu2025-ju,8289405,Jin2025-od,van-Ackooij2013-vt,van-Ackooij2018-ar} &
\cell{0.17}{DIU} &
$\times$ & $\circ$ & $\circ$ & $\circ$ &
\cell{0.21}{Chance-constrained, distributionally robust, SDDP} \\
\refs{Hamann2017-al,su2025real,Cai2025-pcmp,Zhang2025-tg,Yang2025-ece,Lu2025-ece,qi2026online} &
\cell{0.17}{DIU, updated online} &
\checkmark & $\times$ & $\circ$ & $\circ$ &
\cell{0.21}{MPC, reinforcement learning, online convex optimization} \\
\refs{Zhang2022-yy,chen2022robust,qi2023chance,Moreira2024-dr,pianco2025decision} &
\cell{0.17}{DDU, static mapping} &
$\circ$ & $\times$ & $\times$ & $\times$ &
\cell{0.21}{Robust, chance-constrained, distributionally robust} \\
\refs{Zhao2026-qk} &
\cell{0.17}{DDU (generation boundaries)} &
$\times$ & $\times$ & $\times$ & $\circ$ &
\cell{0.21}{Decoupled optimization} \\
\hline
\cell{0.20}{\textbf{This paper}} &
\cell{0.17}{\textbf{DDU (hourly release-dependent inflow)}} &
\stkc{\checkmark\how{hourly MPC}} &
\stkc{\checkmark\how{SSH}} &
\stkc{\checkmark\how{reference tracking}} &
\stkc{\checkmark\how{McCormick}} &
\cell{0.21}{\textbf{Rolling-horizon joint chance-constrained optimization}} \\
\hline
\end{tabular}
{\scriptsize \checkmark: a focus of the cited works; $\circ$: considered in some of the cited works or in simplified form; $\times$: not considered.}
\end{table*}

We organize the remainder of the paper as follows. Section~\ref{opti} presents the dispatch formulation and the DDU model. Section~\ref{solution} introduces the solution algorithm for optimization under DDU. Section~\ref{analysis} presents case studies that validate the framework and benchmark its comparative performance. Section~\ref{conclusion} concludes the paper.

\section{ Formulation and Preliminaries}\label{opti}

We first introduce the streamflow model and its uncertainty characterization, developing a DDU model that captures how upstream releases reshape downstream inflow distributions in real time. We then formulate the real-time cascaded hydropower dispatch as a rolling-horizon joint chance-constrained optimization problem and provide a convex relaxation.

\subsection{Inflow Models and Uncertainty Characterization}\label{uncertainty}

In real-time dispatch, the inflow uncertainty manifests as the forecast error around a deterministic mean prediction issued at each time step. We model the inflow vector $\hat{\boldsymbol{q}}_t = [\hat{q}_{1,t}, \dots, \hat{q}_{n,t}]^\top$ as multivariate Gaussian distribution:
\begin{equation}\label{conditional}
    \hat{\boldsymbol{q}}_t = \boldsymbol{\mu}_t + \boldsymbol{\xi}_t, \quad \boldsymbol{\xi}_t \sim \mathcal{N}(\mathbf{0}, \boldsymbol{\Sigma}_t)
\end{equation}
where $\boldsymbol{\mu}_t \in \mathbb{R}^n$ is the deterministic mean forecast and $\boldsymbol{\xi}_t \in \mathbb{R}^n$ the zero-mean forecast error with covariance $\boldsymbol{\Sigma}_t$. The Gaussian assumption in~\eqref{conditional} is adopted for its accuracy in modeling streamflow forecast errors. Non-Gaussian forecast errors can be accommodated via Gaussian Mixture Models~\cite{zhuang2025real} or other parametric families.

\subsubsection{Mean Forecast Model}

The mean forecast for each unit is learned via an autoregressive model:
\begin{equation}\label{g-ols}
    \mu_{i,t} = \alpha_0 + \alpha_1\, q_{i,t-1} + \beta_{1}\bigl(u_{i-1,t-\tau_i}+s_{i-1,t-\tau_i}\bigr)
\end{equation}
where $i-1$ indexes the upstream unit. The coefficients $\alpha_1$ and $\beta_1$ are estimated offline via least-squares regression. Including past observed inflow $q_{i,t-1}$ captures the dynamics of the local streamflow. Upstream release is included to capture additional flow propagation effects. To estimate the travel time $\tau_i$, we compute the correlation between the total upstream release and downstream inflow
\begin{equation}\label{lag-xcorr}
    \tau_i = \operatorname*{arg\,max}_{h \ge 0}\;\;
   \operatorname{corr} \left( u_{i-1,t-h} + s_{i-1,t-h},\; q_{i,t} \right)
\end{equation}
\noindent to find the hour $\tau_i$ that maximizes correlation.

\subsubsection{Decision-Dependent Uncertainty (DDU) Model}\label{subsec:variance}

\begin{figure}[t]
\centering
\subfloat[\label{fig:residuals}]{%
    \includegraphics[width=0.9\columnwidth]{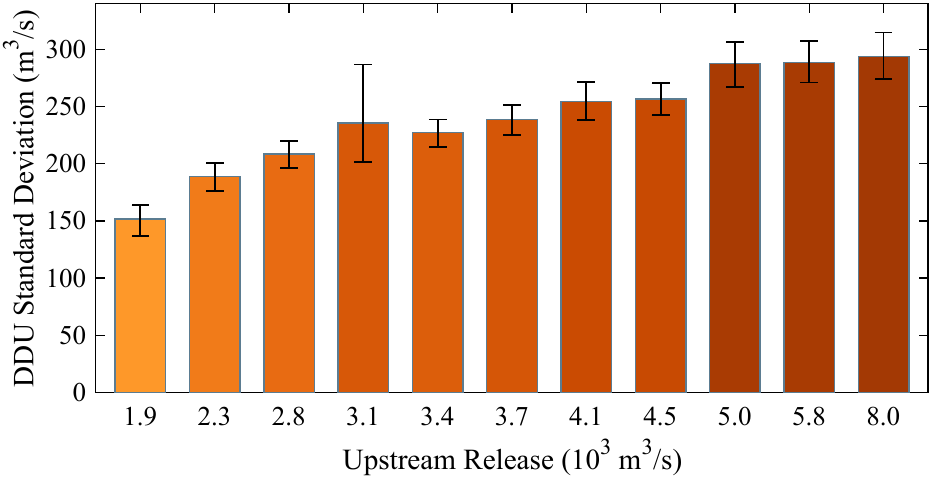}
} \\
\vspace{-1em}
\subfloat[\label{fig:variance}]{%
    \includegraphics[width=0.9\columnwidth]{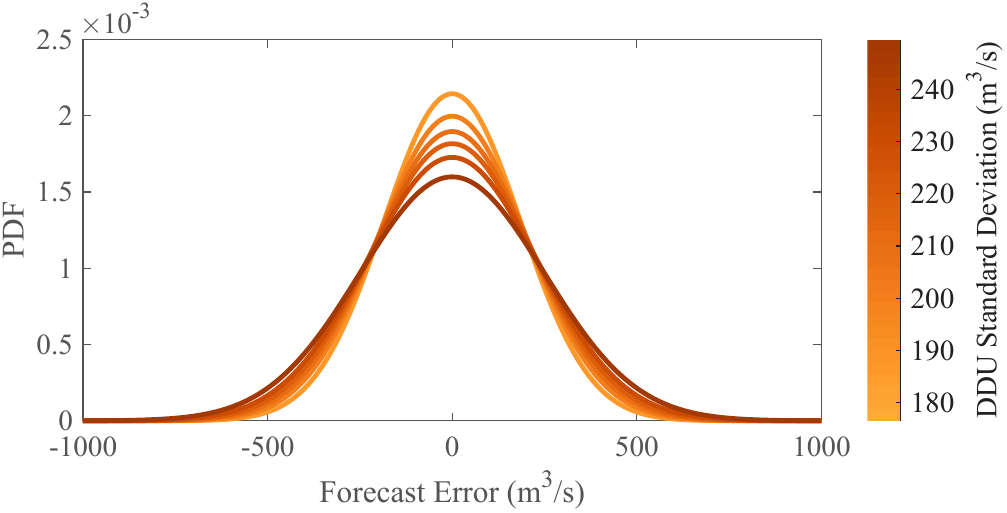}
} \\
\vspace{-1em}
\subfloat[\label{fig:variance-test}]{%
    \includegraphics[width=0.9\columnwidth]{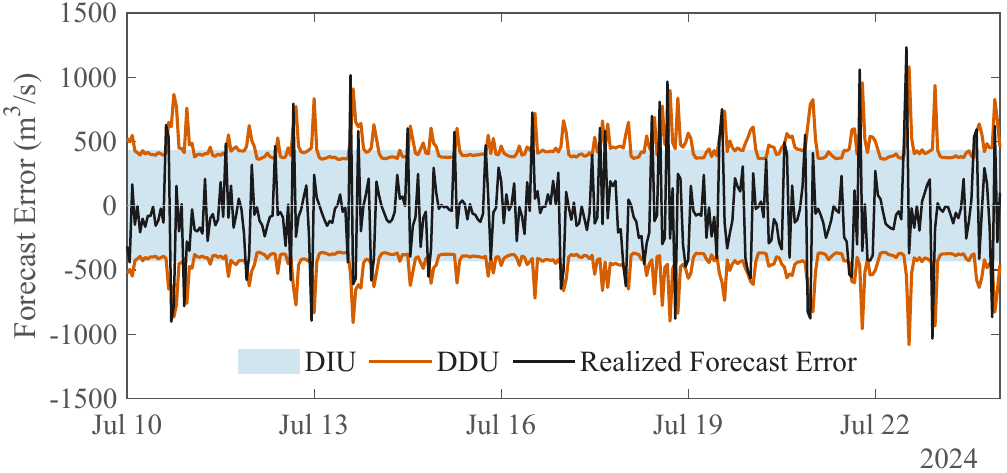}
}
\caption{Empirical evidence for DDU in cascaded hydropower: 
(a) inflow forecast error standard deviation versus upstream release
(b) derived DDU and DIU normal distributions (c) DDU estimation of test data forecast error.}
\label{fig:residual_variance}\vspace{-1em}
\end{figure}

The mean forecast residuals exhibit time-varying volatility associated with upstream releases, consistent with streamflow forecast errors that grow with flow magnitude~\cite{Schoups2010-wrr}. Fig.~\ref{fig:residual_variance} illustrates this on the Lower Columbia River data used in Section~\ref{analysis}: the standard deviation of the forecast error grows with upstream release (Fig.~\ref{fig:residuals}), which widens the error distribution under high releases (Fig.~\ref{fig:variance}), and a release-conditioned variance captures forecast-error spikes on test data (Fig.~\ref{fig:variance-test}). To capture this, we extend the autoregressive conditional heteroskedasticity framework~\cite{Bollerslev1990-tq} with upstream releases as an additional exogenous predictor (ARCH-X). This introduces the endogenous coupling between forecast variance and operational decisions that defines DDU.
\begin{subequations}\label{eq:ddu}
\begin{align}
    &\boldsymbol{\Sigma}_t^{\text{DDU}} = \boldsymbol{D}_t\, \boldsymbol{R}\, \boldsymbol{D}_t \\
    &\boldsymbol{D}_t = \mathrm{diag}(\sigma_{1,t}^{\text{DDU}}, \dots, \sigma_{n,t}^{\text{DDU}}) \\
    &(\sigma_{i,t}^{\text{DDU}})^2 = \omega + \alpha\,e_{i,t-1}^2 + \gamma( u_{j,t-\tau_i}+s_{j,t-\tau_i})\label{g-GARCH-X}
\end{align}
\end{subequations}
where $\boldsymbol{R}$ is the constant correlation matrix with entries $\rho_{ij}$ capturing the spatial correlation of forecast residuals between units $i$ and $j$, and $j = i-1$. The parameters $(\omega, \alpha, \gamma)$ are estimated via maximum likelihood on historical forecast residuals. Note that $\gamma$ quantifies the decision-dependent effect of upstream release on the next-step forecast variance.

By updating $\boldsymbol{\Sigma}_t^{\text{DDU}}$ at each rolling-horizon step, our forecast model couples both the mean $\boldsymbol{\mu}_t$ and covariance $\boldsymbol{\Sigma}_t^{\text{DDU}}$ to past upstream releases, providing a real-time, fully decision-dependent Gaussian representation of inflow uncertainty:
\begin{equation}\label{full-ddu}
    \hat{\boldsymbol{q}}_t \sim \mathcal{N}\!\left(\boldsymbol{\mu}_t(\boldsymbol{u}_{t-\tau} + \boldsymbol{s}_{t-\tau}),\, \boldsymbol{\Sigma}_t^{\text{DDU}}(\boldsymbol{u}_{t-\tau} + \boldsymbol{s}_{t-\tau})\right)
\end{equation}
This formulation extends prior DDU work that conditions only the first moment on decisions~\cite{chen2022robust,qi2023chance,Zhang2022-yy}, and integrates directly into the joint chance-constrained optimization~\eqref{GES}.

ARCH-X is chosen for three main reasons. First, its closed-form conditional variance preserves the analytical tractability required for chance-constrained reformulation. Second, the linear additive structure of~\eqref{g-GARCH-X} provides direct interpretability of how forecast errors and upstream releases jointly shape the next-step variance, with each parameter $(\omega, \alpha, \gamma)$ admitting a clear physical and statistical meaning. Third, the fitted model achieves close to nominal coverage of the 95\% prediction interval (Section~\ref{subsec:calibration}).  More flexible alternatives, such as $k$-nearest neighbor resampling~\cite{Sharma1996-pz} or neural-network forecasts~\cite{Zealand1999-ge}, can capture richer nonlinear dependencies between release decisions and forecast variance. However, these methods typically lack explainable variance expressions and require scenario-based reformulations that increase the computational cost.

\subsection{Real-Time Cascaded Hydropower Dispatch}

\begin{figure}[!t]
    \centering
    \includegraphics[width=0.8\columnwidth]{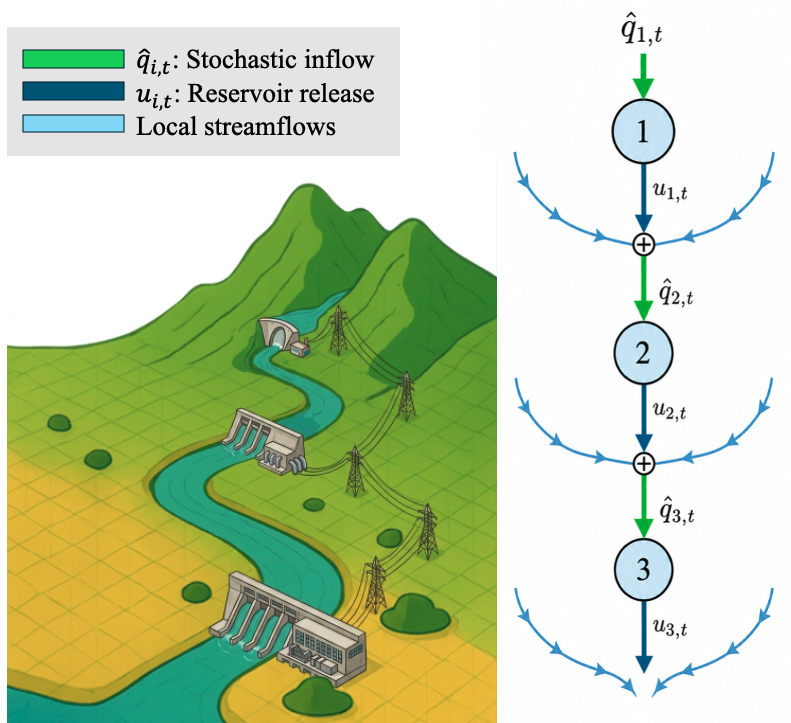}
    \caption{Cascaded hydropower network with correlated streamflows.}
    \label{fig:spatial-flow}\vspace{-1em}
\end{figure}

The real-time joint chance-constrained dispatch problem for a cascaded hydropower system consists of $n$ cascaded units indexed by 
$ \mathcal{I} = \{1\text{,} 2\text{,} \dots\text{,} \ n\}$ is operated over the hours $ \mathcal{T} = \{1\text{,} 2\text{,} \dots\text{,} T\}$, as illustrated in Fig.~\ref{fig:spatial-flow}. Dispatch is set under a rolling horizon such that at each hour $t\in\mathcal{T}$ the problem is solved over the look-ahead window $\mathcal{H}_t=\{t,t+1,\dots,t+H-1\}$. Only the decisions at the first time step are implemented, and the window then advances by one hour and is resolved with updated state and forecasts. A reference trajectory $v^{ref}_{i,t}$ is built from historical data to guide real-time decisions, following~\cite{qi2026online}. The problem is stochastic because the inflow to each reservoir, $\hat{q}_{i,t}$, is stochastic and follows a joint distribution coupled with cascade flow dynamics. The optimization problem is stated as follows ($\forall i \in \mathcal{I}, \forall t \in \mathcal{H}_t$):
\begin{subequations}\label{GES}
\begin{align}
& \max_{p,v,u,s}\; \sum_{t\in\mathcal H_t}\sum_{i\in\mathcal I}\lambda_t p_{i,t}-c_s s_{i,t}-\theta\big|v_{i,t}-v^{ref}_{i,t}\big| \label{g-obj} \\
\text{s.t.}\quad
& v_{i,t} = v_{i,t-1} + \kappa_{\mathrm{af}}(\hat{q}_{i,t} - u_{i,t} - s_{i,t}), \label{g-massbalance} \\
& p_{i,t} = c\,\eta_i g\rho\,h_i(v_{i,t}) u_{i,t}, \label{g-powerfunc} \\
& h_i(v_{i,t}) = (f^{0}_{i} + f_i v_{i,t}) - (t^0_i + t_i\left(u_{i,t-1} + s_{i,t-1}\right)), \label{g-head} \\
& \underline{R}_i \le u_{i,t} - u_{i,t-1} \le \overline{R}_i, \label{g-ramprate} \\
& \underline{U}_i \le u_{i,t} \le \overline{U}_i, \label{g-flowrate} \\
& 0 \le p_{i,t} \le \overline{P}_i, \label{g-feedercap} \\
& \mathbb{P}_{\hat{q}} \left[ \underline{V}_{i,t} \le v_{i,t} \le \overline{V}_{i}, \;\; \forall i \in \mathcal{I} \right] \ge 1-\varepsilon , \label{g-volume} \\
& s_{i,t} \ge 0 \label{g-spill} \\
&\text{\eqref{conditional}-\eqref{g-ols},~\eqref{eq:ddu}}\notag
\end{align}
\end{subequations}

The objective~\eqref{g-obj} maximizes total energy generation revenue across all units and time steps while minimizing deviations from the historical reference trajectory and spill. Let $\lambda_t$ be the price of electricity at time $t$, $c_s$ the cost of spill, and $\theta$ be the penalty term for deviating from the reference trajectory. Constraint~\eqref{g-massbalance} defines the water mass balance equation where turbine release is denoted by $u_{i,t}$, spill as $s_{i,t}$, and stochastic inflow as $\hat{q}_{i,t}$. The inflow random variable $\hat{q}_{i,t}$ is composed of its mean forecast $\mu_{i,t}$, given in~\eqref{g-ols} and error term $\xi_{i,t}$ given in~\eqref{eq:ddu}. The mean forecast is a linear combination of the water released from the upstream hydropower station $i-1$ with $\tau_i$ travel time and $q_{i,t-1}$ is the previously observed local inflow. The nonconvex hydropower production~\eqref{g-powerfunc} calculates power production as a function of reservoir volume and turbine release. Note that~\eqref{g-powerfunc} includes a bilinear term which is relaxed in Section~\ref{mccormick} to ensure tractability.  We define the hydraulic head $h_i(v_{i,t})$ as the difference between the forebay and tailwater elevations, where parameters $(f^0_i, \; f_i, \; t^0_i, \; t_i)$ are fit from real-world data, following~\cite{Diniz2008-fd}. Tailwater is evaluated at $t-1$ since previous conditions are known to the operator. The parameters $\eta$, $g$, $\rho$ denote the turbine efficiency, gravitational acceleration, and water density, respectively. $c=10^{-6}$ denotes the conversion factor from watts to megawatts.  Constraints~\eqref{g-ramprate}--\eqref{g-volume} impose the ramping, water release, generation, and volume limits, respectively where $[\underline{R}_i, \overline{R}_i]$, $[\underline{U}_i, \overline{U}_i]$, $\overline{P}_i$, $[\underline{V}_i, \overline{V}_i]$ denote the plant-specific bounds. To maintain reservoir volume safety under uncertain inflows, the joint chance constraint~\eqref{g-volume} ensures all reservoir volume bounds are simultaneously satisfied across units for each time step with $1-\varepsilon$ confidence level. Constraint~\eqref{g-spill} ensures spill is positive at all time steps. We treat the cascade as a regulated price-taker following the current case in Bonneville Power Administration (BPA), where generation is sold to customers at FERC approved tariff rates rather than into a wholesale market~\cite{Bonneville_Power_Administration2022-ez}. The formulation extends directly to market prices by updating $\lambda_t$ with hourly forecasts in the rolling horizon, since the chance constraints and the SSH algorithm are unaffected; price uncertainty and its correlation with inflow are left for future work.

\subsection{McCormick Relaxation for Hydropower Production}\label{mccormick}

The hydropower generation function exhibits non-convexity due to the bilinear dependence of hydraulic head $h(v_{i,t})$ and turbine release $u_{i,t}$. To address this, the McCormick approximation~\cite{bynum2018tightening} is employed to replace the bilinear term in~\eqref{g-powerfunc} with a convex counterpart: 

\begin{subequations}\label{m-env}
    \begin{align}
        & p_{i,t} \geq c\,\eta_i g\rho\, (\underline{h}_iu_{i,t} + h_{i,t}\underline{u}_i - \underline{h}_i\underline{u}_i)\label{under}\\
        & p_{i,t} \geq c\,\eta_i g\rho\, (\overline{h}_iu_{i,t} + h_{i,t}\overline{u}_i - \overline{h}_i\overline{u}_i) \\
        & p_{i,t} \leq c\,\eta_i g\rho\, (\overline{h}_iu_{i,t} + h_{i,t}\underline{u}_i - \overline{h}_i\underline{u}_i) \\
        & p_{i,t} \leq c\,\eta_i g\rho\, (h_{i,t}\overline{u}_{i} + \underline{h}_{i}u_{i,t} - \underline{h}_i\overline{u}_i) \label{upper}
    \end{align}
\end{subequations}

This linearization is a standard method for relaxing bilinear terms in hydropower operations with error analysis in~\cite{bynum2018tightening}.

\section{Solution Methodology}\label{solution}

The optimal dispatch policy for the joint chance-constrained dispatch problem~\eqref{GES} under DDU is obtained using the proposed Sequential Supporting Hyperplane (SSH) method. We present the algorithm, establish its convergence to the global optimum, and characterize its adaptive risk allocation property. 

\subsection{Sequential Supporting Hyperplane Algorithm}

\begin{figure}[!t]
    \centering
    \includegraphics[width=0.90\columnwidth]{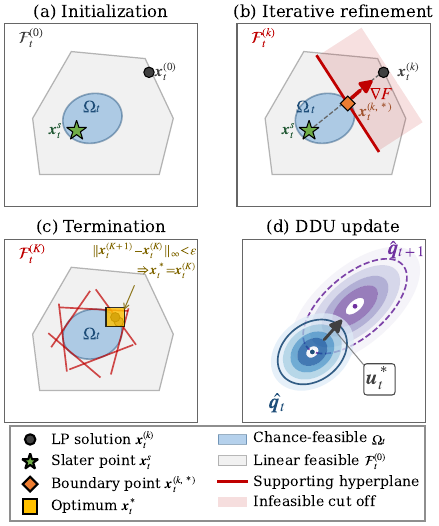}
    \caption{Schematic of the proposed SSH algorithm.}
    \label{fig:SSH}
\end{figure}
Solving joint chance-constrained optimization under DDU over time is technically hard. Inspired by the literature~\cite{qi2023chance, Zhang2022-yy}, the most effective way is to decouple the decision and uncertainty with iterative algorithm while guaranteeing optimality and convergence performance. In this paper, we propose the SSH algorithm which computes the optimal policy for problem~\eqref{GES} by solving for $\boldsymbol{x}_t = (\boldsymbol{u}_t, \boldsymbol{s}_t, \boldsymbol{p}_t)$, as illustrated in Fig.~\ref{fig:SSH}. The method preserves the joint chance constraint by iteratively refining a polyhedral outer approximation of the feasible region using supporting hyperplanes. This algorithm is needed because the joint chance constraint is not directly linearizable without decomposition. Since the inflow vector is modeled as a multivariate Gaussian random vector, the joint chance constraint has a Gaussian rectangle-probability representation. In this context, rectangle denotes the component wise interval $[a,b] =\{x:a_i\leq x_i\leq b_i,\ \forall i\}$. For $X \sim \mathcal{N}(\mu, \Sigma)$ and $a, b \in \mathbb{R}^n$. The explicit integral form of the CDF is:

\begin{equation}\label{CDF}
    F_X(a, b) = \mathbb{P}(a \leq X \leq b) = \int_{a_1}^{b_1} \cdots \int_{a_n}^{b_n} f_X(t)dt
\end{equation}

Next, we reformulate~\eqref{g-volume} as $\mathbb{P}[\underline{V}' \leq  \hat{q}_t \leq \overline{V}'] \geq 1 - \varepsilon$, where $\underline{V}' = \frac{1}{\kappa_{\mathrm{af}}}(\underline{V} - v_{t-1}) + u_t + s_t$ and $\overline{V}' = \frac{1}{\kappa_{\mathrm{af}}}(\overline{V} - v_{t-1}) + u_t + s_t$. Combining~\eqref{CDF} with the reformulated~\eqref{g-volume} yields the probability that the volume bounds are jointly satisfied. 

\begin{equation}\label{joint-dist}
    F(\underline{V}', \overline{V}') = \frac{1}{(2\pi)^{\frac{n}{2}} |\hat{\Sigma}_t|^{\frac{1}{2}}} \int^{\overline{V}'_{1}}_{\underline{V}'_{1}} \cdots \int^{\overline{V}'_{n}}_{\underline{V}'_{n}} e^{-\frac{1}{2}(x - \mu)^\top \hat{\Sigma}_t^{-1} (x - \mu)}dx
\end{equation}

At each time step, the $n$-dimensional Gaussian rectangle probability in~\eqref{joint-dist} using Genz's code~\cite{Genz1992-gs}, and its gradient is computed by finite differences. Therefore, horizon length increases the number of evaluations but not the CDF dimension. We note the probability under DDU depends on system decisions and is referred to as $F(\underline{V}'(\boldsymbol{x}_t), \overline{V}'(\boldsymbol{x}_t)$. Because the multivariate Gaussian CDF is log-concave, the feasible region is convex and can therefore be approximated using supporting hyperplanes. 

We now introduce the algorithm which consists of four components: 1) initialization, 2) iterative refinement through cutting-planes, 3) termination, and 4) DDU updates.

\begin{enumerate}
    \item \textit{Initialization: } At each time step $t$:

\begin{enumerate}[a)]
    \item Solve~\eqref{GES} without the chance constraint~\eqref{g-volume} to obtain $\boldsymbol{x}_t^{(0)}$. This solution maximizes generation under the deterministic constraints~\eqref{g-massbalance}--\eqref{g-feedercap}, but may violate the chance constraint~\eqref{g-volume}.
    
    \item Construct a strictly feasible point $\boldsymbol{x}_t^s = (\boldsymbol{u}_t^s, \boldsymbol{s}_t^s, \boldsymbol{p}_t^s)$ satisfying~\eqref{joint-dist}. This point keeps reservoir volumes within admissible bounds and is defined using a conservative minimum-release policy:
        \begin{subequations}\label{slater}
        \begin{align}
            \boldsymbol{u}_t^s &= \max\{\underline{\boldsymbol{U}}, \boldsymbol{u}_{t-1} - \underline{\boldsymbol{R}}\} \\
            \boldsymbol{s}_t^s &= \max\Big\{\boldsymbol{0},\; \frac{1}{\kappa_{\mathrm{af}}}\big(\boldsymbol{v}_{t-1} - \overline{\boldsymbol{V}}\big) + \hat{\boldsymbol{q}}_t - \boldsymbol{u}_t^s \Big\} \\
            \boldsymbol{v}_t^s &= \boldsymbol{v}_{t-1} + \kappa_{\mathrm{af}}\big(\hat{\boldsymbol{q}}_t - \boldsymbol{u}_t^s - \boldsymbol{s}_t^s\big) \\
            \boldsymbol{p}_t^s &= c\,\eta\,g\,\rho\,h(\boldsymbol{v}^s_{t})  \boldsymbol{u}_t^s
        \end{align}
        \end{subequations} 
        
    \item Initialize the constraint matrices $(\boldsymbol{A}_t^{(0)}, \boldsymbol{b}_t^{(0)}) \gets (\boldsymbol{A}_t, \boldsymbol{b}_t)$, where $\mathbf{A}_t, \mathbf{b}_t$ denote deterministic coefficient matrices and vectors that encode constraints~\eqref{g-massbalance}--\eqref{g-feedercap}. 
\end{enumerate}
\item 
    \textit{Iterative Refinement: } At each iteration $k$:
    \begin{enumerate}[a)]
        \item Compute a convex combination between the current solution $\boldsymbol{x}_t^{(k)}$ and the feasible point $\boldsymbol{x}_t^s$
        \[
            \boldsymbol{x}_t^{(k,*)} = (1-\lambda^*)\boldsymbol{x}_t^{(k)} + \lambda^* \boldsymbol{x}_t^s
        \]
        where $\lambda^* \in (0,1]$ is computed by a bisection line search such that $\boldsymbol{x}_t^{(k,*)}$ lies on the boundary of the chance constrained feasible region defined by $F(\boldsymbol{x}_t^{(k,*)}) = 1 - \varepsilon$.  
        
        \item Evaluate the gradient $\nabla F$ at $\boldsymbol{x}_t^{(k,*)}$. Since the feasible region $\{\boldsymbol{x}_t \mid F(\boldsymbol{x}_t) \ge 1 - \varepsilon\}$ is a superlevel set of $F$, the linear constraint 
        \[
            \nabla F^\top(\boldsymbol{x}_t - \boldsymbol{x}_t^{(k,*)}) \ge 0
        \]
        defines a supporting hyperplane at $\boldsymbol{x}_t^{(k,*)}$.
    
        \item Append the new hyperplane to the constraint set:
\begin{equation*}
\boldsymbol{A}_t^{(k+1)} = 
\begin{bmatrix}
\boldsymbol{A}_t^{(k)} \\
\nabla F^\top
\end{bmatrix}
\;\;, \quad\;\;
\boldsymbol{b}_t^{(k+1)} = 
\begin{bmatrix}
\boldsymbol{b}_t^{(k)} \\
\nabla F^\top \boldsymbol{x}_t^{(k,*)}
\end{bmatrix}
\end{equation*}
    
        which yields the refined feasible region $\mathcal{F}_t^{(k+1)}$. 
        
        \item Solve the linear program over the updated region: 
        \[
            \boldsymbol{x}_t^{(k+1)} = \arg\max \left\{  \boldsymbol{p}_t \;\middle|\; \boldsymbol{A}_t^{(k+1)} \boldsymbol{x}_t \le \boldsymbol{b}_t^{(k+1)} \right\}
        \]
    \end{enumerate}
\item 
    \textit{Termination:} The iterations stop when consecutive solutions satisfy 
    \[
    \|\boldsymbol{x}_t^{(k+1)} - \boldsymbol{x}_t^{(k)}\|_\infty < \epsilon
    \]
\item \textit{DDU Updates:} The final solution is set to $\boldsymbol{x}_t^* = \boldsymbol{x}_t^{(k)}$. The covariance $\boldsymbol{\Sigma}^{\mathrm{DDU}}$ is updated via~\eqref{g-GARCH-X} using the implemented release $\boldsymbol{u}_t^*$, thereby propagating decision dependence to the next step. 
\end{enumerate}

\subsection{Theoretical Analysis}

\begin{theorem}[Convergence]\label{the1}
The SSH algorithm terminates after a finite number of iterations $K$ with $\|\boldsymbol{x}_t^{(K+1)} - \boldsymbol{x}_t^{(K)}\|_\infty < \epsilon$ and returns $\boldsymbol{x}_t^* = \boldsymbol{x}_t^{(K)}$ as an $\epsilon$-optimal solution to~\eqref{GES}.
\end{theorem}

Theorem~\ref{the1} establishes the key algorithmic guarantee: for any prescribed tolerance $\epsilon$, the SSH algorithm is guaranteed to terminate within a finite number of iterations with an $\epsilon$-optimal solution. The solution convergence to the algorithm relies on two structural properties (a) the chance-constrained feasible region $\Omega_t$ is convex due to the log-concavity of the multivariate Gaussian CDF, and (b) the sequence of supporting hyperplane cuts constructs nested polyhedral outer approximations $\mathcal{F}_t^{(k)}$ that monotonically shrink toward $\Omega_t$. The Slater point construction~\eqref{slater} provides the strictly feasible interior point required for both the line-search step in the algorithm and the convergence proof. This guarantee makes SSH suitable for real-time dispatch with rolling-horizon updates, where bounded computation time at each step is essential. The proof is provided in Appendix~\ref{thm-convergence}.  SSH extends to convex MINLP formulations but uses linear subproblems here for real-time efficiency. General nonconvex models would sacrifice its global convergence guarantee~\cite{Kronqvist2016-ai}. 

\begin{proposition}[Risk allocation under steady-state]\label{prop1}
Under steady-state conditions, the SSH risk allocation is equally distributed across units, i.e., $\lim_{t\to\infty}\varepsilon^{\mathrm{SSH}}_{i,t} \rightarrow \frac{\varepsilon}{n}$.
\end{proposition}

Symmetric steady-state refers to conditions where releases vary negligibly over time and the uncertainty
distributions and operating bounds are symmetric across units. The proof of the adaptive risk allocation is provided in Appendix~\ref{thm-risk}.

Proposition~\ref{prop1} establishes a consistency property: in equilibrium with symmetric operating conditions, the SSH risk allocation reduces to the uniform allocation $\varepsilon/2n$ used by classical Bonferroni-based approximations~\cite{Yang2022-tw}. However, cascaded hydropower systems rarely operate at steady state in practice. Streamflow exhibits persistent non-stationarity driven by seasonal hydrologic cycles, weather variability, and increasingly frequent climate extremes such as droughts and floods. Under these transient regimes, units face heterogeneous risk exposures: downstream units experience amplified inflow uncertainty following upstream releases, and constrained units approach their volume bounds at different rates. Fixed uniform allocations become overly conservative for units operating in the interior and insufficiently protective for units near the boundary. SSH addresses this by dynamically allocating risk based on the CDF gradient or allocating uniform risk under the symmetric steady-state conditions, as shown in  Proposition~\ref{prop1}. This adaptability is particularly valuable for real-time dispatch under climate variability, where transient disturbances dominate operational performance.

\subsection{Bonferroni Solution Algorithm}

The Bonferroni approximation provides a benchmark for the SSH algorithm by decomposing the joint chance constraint into \textit{individual chance constraints}. Each individual constraint is reformulated exactly by its Gaussian quantile, distributing the total risk budget $\varepsilon$ uniformly across units such that $\varepsilon_i^{BON} = {\varepsilon}/{2n}$ for each time step. The complete derivation is given in Appendix~\ref{app:bon}.

\section{Numerical Studies}\label{analysis}

\subsection{Setup}

We consider a cascaded hydropower system that models the four lower
hydropower reservoirs in the Columbia River Basin, which listed in cascade
order are the McNary, John Day, Dalles, and Bonneville Dam. The nameplate
capacity of the units are 980, 2160, 1878, and 1227 MW respectively, and
each are modeled with an 88\% generating efficiency.  The minimum and maximum
turbine flow rates for each reservoir are [1274, 6612],
[1266, 9333], [1076, 8293], and [824, 7377] m$^3$/s. The ramp-up and
ramp-down rates are [-1167, 1201], [-1555, 1587], [-1376, 1387], and
[-1076, 1011] m$^3$/s/hr, respectively. The real-time dispatch problem is solved in
receding horizon over a 24-hour look-ahead window across the full 2024 year of 8760 hours.
The Columbia River Basin system and operating data is available
online.\footnote{Data Available: https://www.nwd.usace.army.mil/columbia-river/}

This case study simulates 2024 water year (WY2024) from October 2023 to September 2024, capturing low-flow periods during the fall and high flow periods during the summer. This setting makes system performance highly sensitive to inflow uncertainty because forecast errors directly increase the risk of constraint violations. All case studies are
coded in MATLAB and solved by Gurobi 12.02 solver. The programming environment is Apple M3 4.1 GHz CPU and 16 GB of RAM\footnote{The code and data used in this study is available at: https://github.com/ecohn44/cascaded-hydro}.

\subsection{Inflow Model Learning}\label{subsec:calibration}

\begin{figure}[t]
    \centering
    \includegraphics[width=0.9\columnwidth]{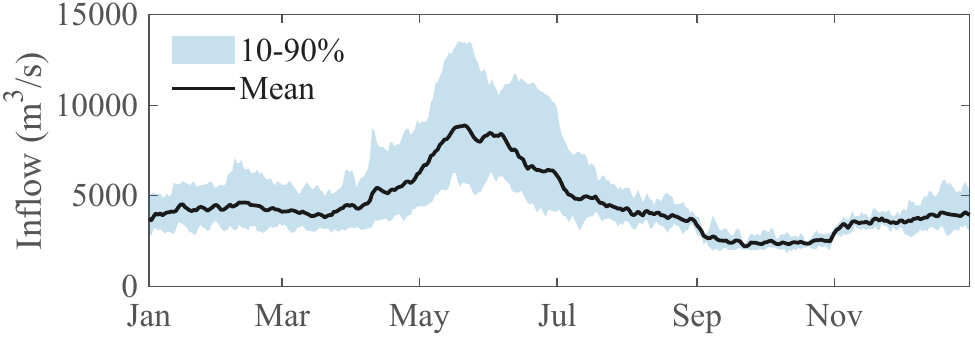}
    \caption{Seasonal streamflow hydrograph for the Lower Columbia River.}
    \label{fig:hydrograph}\vspace{-1em}
\end{figure}

We fit the inflow models of Section~\ref{uncertainty} to hourly records of the four projects. The ARCH-X parameters are estimated on eight water years of residuals from 2018--2025, which include $70{,}080$ hourly observations per project. Fig.~\ref{fig:hydrograph} shows the seasonal hydrograph of the Lower Columbia River, with low flows in the fall and high flows in the summer.

Table~\ref{tab:meanmodel} lists the normalized fits of the mean model~\eqref{g-ols}. Autoregressive coefficients $\alpha_1$ close to one indicate strong hourly inflow persistence that benefits from real-time state updates, and the estimated travel times are 10, 3, and 4 hours for John Day, The Dalles, and Bonneville. Although the mean model explains most of the inflow variation ($R^2 \geq 0.96$), the remaining residual spread varies with upstream release, as shown in Fig.~\ref{fig:residual_variance}.

\begin{table}[t]
\centering
\caption{Mean Streamflow Model Normalized Fits}
\label{tab:meanmodel}
\renewcommand{\arraystretch}{0.85}
\begin{tabular*}{0.9\columnwidth}{@{\extracolsep{\fill}}clrrrrr@{}}
\toprule
Unit & Dam & $\alpha_0$ & $\alpha_1$ & $\beta_1$ & $\tau$ (hr) & $R^2$ \\
\midrule
1 & McNary$^\ast$ & 0.000 & 1.000 &  0.000 & -- & 0.981 \\
2 & John Day      & 0.426 & 0.929 &  0.069 & 10 & 0.974 \\
3 & The Dalles    & 1.369 & 1.037 & -0.044 &  3 & 0.964 \\
4 & Bonneville    & 0.401 & 1.006 & -0.009 &  4 & 0.993 \\
\bottomrule
\end{tabular*}
\vspace{1pt}

{\footnotesize $^\ast$Upstream unit uses local inflow.}
\end{table}

Table~\ref{tab:archx} lists the normalized ARCH-X fits of~\eqref{g-GARCH-X}. The release coefficient $\gamma$ is positive for the three downstream units and zero for McNary, which uses local inflow. Goodness of fit is measured by cov$_{95}$, the fraction of standardized residuals $e_t/\sigma_t$ falling within $\pm 1.96$; the fitted models cover 92--95\% of the residuals against the nominal 95\%.

\begin{table}[t]
\centering
\caption{ARCH-X conditional-variance normalized fits}
\label{tab:archx}
\renewcommand{\arraystretch}{0.85}
\begin{tabular*}{0.9\columnwidth}{@{\extracolsep{\fill}}lrrrr@{}}
\toprule
Dam & $\omega$ & $\alpha$ & $\gamma$ & cov$_{95}$ \\
\midrule
McNary     & 0.632 & 0.441 &  0.00 & 0.931  \\
John Day   & 0.447 & 0.239 &  5.17 & 0.922  \\
The Dalles & 0.100 & 0.061 & 10.73 & 0.918 \\
Bonneville & 0.190 & 0.352 &  8.79 & 0.945 \\
\bottomrule
\end{tabular*}
\end{table}

The DIU benchmark follows the common practice of characterizing the residual covariance as a constant matrix estimated offline from historical forecast residuals:
\begin{subequations}\label{forecast-error}
\begin{align}
    \boldsymbol{\Sigma}^{\text{DIU}} &= \frac{1}{T_h - 1} \sum_{t=1}^{T_h} (\boldsymbol{e}_t - \bar{\boldsymbol{e}})(\boldsymbol{e}_t - \bar{\boldsymbol{e}})^\top \label{sigma-diu-mat} \\
    \boldsymbol{\sigma}^{\text{DIU}} &= \sqrt{\operatorname{diag}\!\left(\boldsymbol{\Sigma}^{\text{DIU}}\right)} \label{sigma-diu}
\end{align}
\end{subequations}
where $\boldsymbol{e}_t = (e_{1,t}, \dots, e_{n,t})^\top$ is the historical residual vector, $T_h$ the sample size, and $\bar{\boldsymbol{e}}$ the empirical mean. The resulting per-unit standard deviations are $\sigma^{\mathrm{DIU}}_i = [397.1,\,218.6,\,226.4,\,238.3]$~m$^3$/s, held constant at each hour. This fixed covariance can underestimate forecast risk during high-release periods and impose excessive conservatism during low-release periods.

\subsection{Policy Testing under Stochastic Streamflows}

\begin{figure}[!t]
    \centering
    \includegraphics[width=0.95\columnwidth]{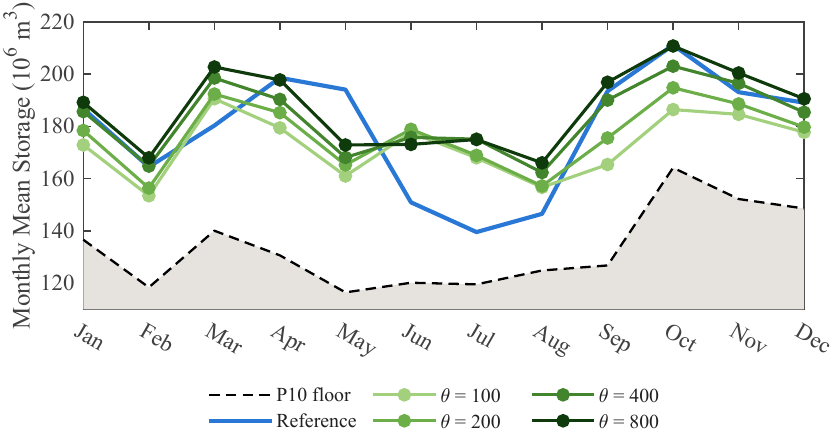}
    \caption{Optimal policy trajectories under DDU for the monthly average volumes. The blue trajectory is the reference, the dashed line is the $P_{10}$ lower envelope, and the shaded region below is for violations.}
    \label{fig:tracking}
\end{figure}

\begin{table}[t]
\centering
\footnotesize
\caption{DDU vs DIU Generation, Violations, Revenue, and Penalties under SSH
for $\theta=400$ over error tolerance $\varepsilon$.}
\label{tab:ssh_results}
\setlength{\tabcolsep}{8.5pt}
\begin{tabular}{@{}cccccc@{}}
\toprule
$\varepsilon$ & Model & Generation & IVI & Revenue & Penalty \\
 & & (GWh) & ($10^6$ m$^3$) & (M\$) & (k\$) \\
\midrule
0.01 & DIU & 26,606.1 & 1,707.1 & 1,032.8 & 567 \\
     & DDU & 26,618.3 & 1,137.3 & 1,033.6 & 291 \\
\midrule
0.05 & DIU & 26,597.9 & 2,111.7 & 1,032.2 & 919 \\
     & DDU & 26,608.9 & 1,650.4 & 1,033.0 & 554 \\
\midrule
0.10 & DIU & 26,594.5 & 2,247.4 & 1,031.9 & 1,116 \\
     & DDU & 26,602.4 & 1,846.5 & 1,032.6 & 778 \\
\midrule
0.15 & DIU & 26,592.7 & 2,626.1 & 1,031.8 & 1,448 \\
     & DDU & 26,601.9 & 1,942.7 & 1,032.5 & 941 \\
\midrule
0.20 & DIU & 26,589.6 & 2,631.0 & 1,031.6 & 1,657 \\
     & DDU & 26,597.3 & 2,142.6 & 1,032.2 & 1,188 \\
\midrule
0.25 & DIU & 26,588.2 & 2,742.0 & 1,031.5 & 1,826 \\
     & DDU & 26,596.8 & 2,215.3 & 1,032.1 & 1,452 \\
\bottomrule
\end{tabular}
\end{table}

To compare the performance of different uncertainty frameworks $m$, we
perform Monte Carlo simulations under stochastic inflow conditions. System reliability is quantified using the integrated violation index (IVI), which measures the cumulative reservoir volume violations over the simulation horizon. $C^{(m)}$ is the associated cost of the constraint violation. 

\begin{subequations}
\begin{align}
& IVI^{(m)} = \frac{1}{S} \sum_{s = 1}^{S} \sum_{i = 1}^{N} \sum_{t = 1}^{T}
\max\!\left(\underbar{V}_{i,t} - v_{i,t}^{(s)},\, 0\right) \label{IVI} \\
& C^{(m)} = \frac{1}{S} \sum_{s=1}^{S} \sum_{i=1}^{N} \sum_{t=1}^{T}
\lambda_t \, c\,\eta_i\, h_{i,t}^{(s)} \, \Delta u_{i,t}^{(s)} \label{cost}
\end{align}
\end{subequations}

\noindent where $\underbar{V}_{i,t}$ is the $P_{10}$ envelope and $v_{i,t}^{(s)}$ is the storage realized after recourse. For each Monte Carlo scenario $s = 1,\dots,S$:

\begin{enumerate}
    \item Sample the forecast errors
    $e_{i,t}^{(s)} \sim \mathcal{N}(0,\Sigma_t)$

    \item Construct the stochastic inflow trajectory
    $q_{i,t}^{(s)} = \mu_{i,t} + e_{i,t}^{(s)}$ 
    
    \item Simulate
    $\tilde{v}_{i,t}^{(s)} = v_{i,t-1}^{(s)} + \kappa_{\mathrm{af}}
    (q_{i,t}^{(s)} - u_{i,t}^{(m)*} - s_{i,t}^{(m)*})$.

    \item Apply recourse to $\tilde{v}$ if the realized trajectory is infeasible by reducing release ($\Delta u_{i,t}^{(s)}$) or increasing spill.

    \item Record the total energy production, constraint violations IVI~\eqref{IVI} and cost of curtailment~\eqref{cost} under the sampled
    trajectory.
\end{enumerate}

\begin{figure*}[t]
\centering
\subfloat[\label{fig:diff-policy-heatmap-energy}]{%
    \includegraphics[width=0.45\textwidth]{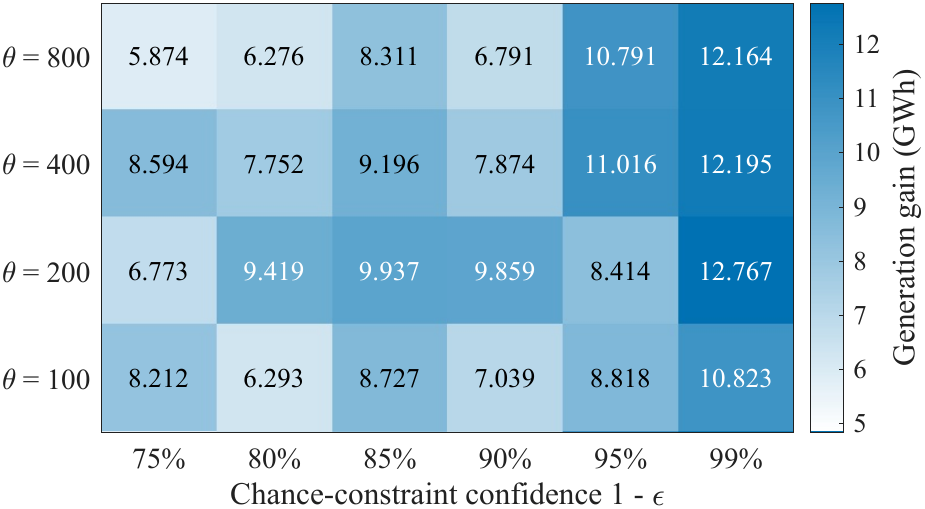}
}
\hfill
\subfloat[\label{fig:diff-policy-heatmap-ivi}]{%
    \includegraphics[width=0.45\textwidth]{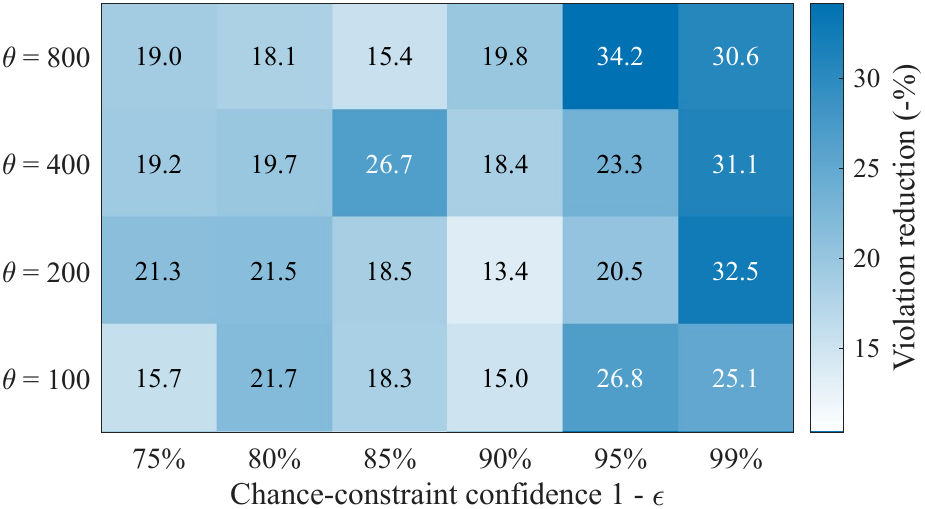}
}
\caption{Improvement of DDU relative to DIU across tracking penalty $\theta$ and uncertainty attitude $\epsilon$. (a) Increase in total generation (GWh); (b) reduction in integrated violation index (IVI, \%).}
\label{fig:diff-policy-heatmap}
\end{figure*}

Table~\ref{tab:ssh_results} summarizes the policy performance over the full water year under the paired inflow scenarios. DDU achieves the highest cumulative generation because it maintains the highest forebay elevations during steady state operation, resulting in improved generation
efficiency. Solve time for each 24-hour horizon ranged from 175 to 194 (ms). We show that DDU earns more revenue at every $\varepsilon$ and forgoes less generation to recourse release cuts, proving the reliability it gains is not
paid for in output. This result highlights how decision-dependent uncertainty implicitly values stored water and promotes water conservation within a real-time dispatch framework. Fig.~\ref{fig:diff-policy-heatmap} extends the comparison across tracking weights $\theta$ and confidence levels $1-\varepsilon$: DDU increases generation by 5.9--12.8~GWh and reduces IVI by 13.4--34.2\% in every setting. Fig.~\ref{fig:tracking} shows the impact of the tracking
weight $\theta$. We select $\theta=400$ to balance volume protection
and dispatch flexibility. Sensitivity analyses use $\varepsilon=0.05$
to provide a nominal 95\% joint reliability guarantee, consistent
with a conventional operational risk tolerance.

\begin{figure}[t]
    \centering
\includegraphics[width=0.9\columnwidth]{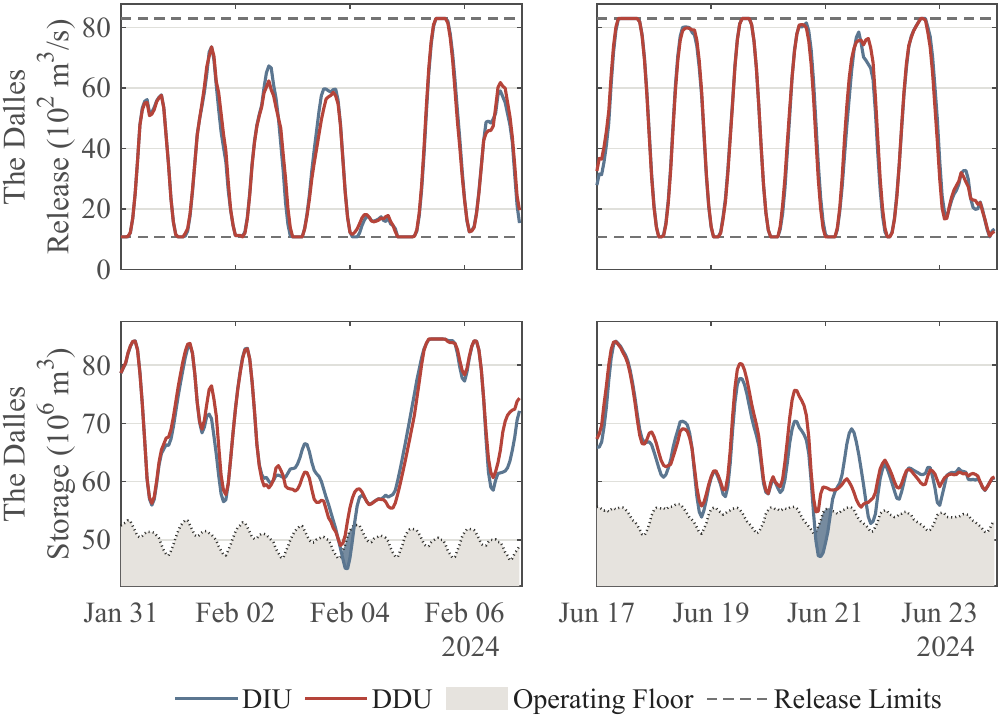}
    \caption{Sample optimal policy trajectories for water release and reservoir state at The Dalles Reservoir under DIU and DDU uncertainty frameworks using the supporting hyperplane solution algorithm with $\varepsilon = 0.05$ and $\theta = 400$.}
    \label{fig:storage}
\end{figure}

Fig.~\ref{fig:storage} illustrates the optimal release policies realized over two weeks in different seasons. Because energy is valued at the BPA load-shaping rate, release tracks the price signal until storage becomes constrained. Under DIU, the margin carried above the floor is fixed regardless of operation, so the policy draws the pool down to the operating floor and crosses below it when a disruption arrives. In contrast, the DDU policy responds more conservatively by reducing release earlier and maintaining higher forebay elevations, providing additional storage margins, remaining above the floor in both windows.

\subsection{Benchmarking Uncertainty Formulations}

\begin{table}[t]
\centering
\caption{Real-time dispatch benchmarks for $\theta=400$ and
$\varepsilon=0.05$ averaged over inflow scenarios.}
\label{tab:bench}
\footnotesize
\setlength{\tabcolsep}{5.5pt}
\begin{tabular}{@{}lcccc@{}}
\toprule
Formulation & $z$ & Energy & IVI & Solve \\
 & & (GWh) & ($10^6$ m$^3$) & (ms) \\
\midrule
Deterministic             & $0$     & $26,567.0$ & $4,428$ & $42.6$ \\
Robust                    & $2.234$ & $26,599.3$ & $1,815$ & $42.8$ \\
Scenario-based            & $2.261$ & $26,599.7$ & $1,922$ & $35.9$ \\
Chance-constrained (DIU)  & $2.241$ & $26,600.2$ & $1,771$ & $31.0$ \\
Chance-constrained (DDU)  & $2.241$ & $26,608.6$ & $1,365$ & $31.4$ \\
\bottomrule
\end{tabular}
\end{table}
The proposed approach is benchmarked against well-established real-time dispatch strategies widely adopted in both the literature and industry~\cite{Roald2023-ng}. Let $\xi_{i,t}$ denote the inflow forecast error at unit $i$ and time $t$, and $\zeta_i=\xi_{i,t}/S_{i,t}$ is its standardized form, where $S_{i,t}=\kappa_{\mathrm{af}}\,\sigma_{i,t}$ be the standard deviation of the volume forecast error in m$^3$. Each formulation enforces the volume constraint differently:

\begin{subequations}
\begin{align}
\underline V_{i} \leq v_{i,t} \leq \overline{V}_{i} & \quad  \text{deterministic} \label{eq:ce}\\
\underline V_{i} \leq v_{i,t}+S_{i,t}\zeta_i \leq \overline{V}_{i} & \quad \text{robust} \;\;(\forall\zeta\in\Xi) \label{eq:ro}\\
\underline V_{i} \leq v_{i,t}+S_{i,t}Z^{(s)}_i \leq \overline{V}_{i} & \quad \text{scenario-based} \;\; (\forall s\in\mathcal S) \label{eq:sp}\\
\mathbb{P}_{\xi} \left[ \underline{V}_{i} \le v_{i,t} \le \overline{V}_{i}  \right] \ge 1-\varepsilon 
 & \quad \text{joint chance constr.} \label{eq:cc}
\end{align}
\end{subequations}

The deterministic formulation ignores uncertainty and has no explicit safety buffer in the volume constraints. Robust optimization protects against every realization $\zeta\in\Xi$, where $\Xi$ is the robust set. Scenario-based optimization considers a finite sample of uncertainty realizations $\mathcal{S}=\{\xi^{(s)}\}$, where $Z^{(s)}_i$ contains the draws of $\zeta_i$. Chance-constrained optimization models $\xi$ directly by its probability distribution and permits constraint violations within violation budget $\varepsilon$.

Table~\ref{tab:bench} reports benchmark performance and mean solution
time per 24-hour horizon. We set each method to guarantee the same reliability, $1-\varepsilon = 0.95$. For robust optimization this requires $\Phi(\Gamma)^{n} = 0.95$ and $\Gamma = 2.234$, which was determined from sweeping $\Gamma$ from $1$ to $3$. For scenario-based optimization we set $|\mathcal S| = 10^3$, the point at which redrawing the sample changes the multiplier $z$ by less than $0.05$. The multipliers $z$ differ slightly because each method reaches the target differently, delivering $0.950$, $0.951$ and $0.953$. Note that the Bonferroni approximation was used for exact control of $z$.

The benchmarks carry the constant DIU forecast variance, since robust and
scenario-based optimization fix their sets and samples before the release
schedule is known. Decision-dependent robust and scenario formulations
also exist~\cite{Zhang2022-yy}. However, their set size or sample count does not directly specify reliability. The proposed chance constraint allows operators to set this requirement explicitly.

\subsection{Sensitivity Analysis}

\subsubsection{Benchmarking Solution Algorithms}

\begin{figure}[t]
    \centering
\includegraphics[width=0.9\columnwidth]{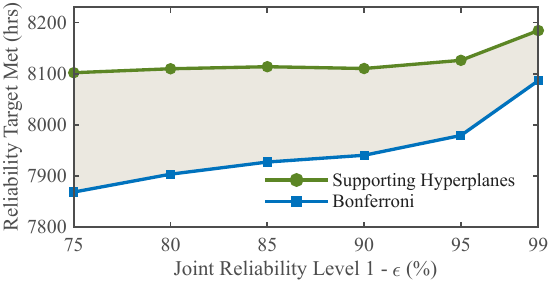}
    \caption{Hours at which the joint reliability target is met over the full year simulation (T = 8760), under the
    Bonferroni approximation and supporting hyperplanes algorithm. The results are averaged over scenarios and $\theta=400$.}
    \label{fig:neighbors}
\end{figure}

The sequential supporting hyperplanes algorithm (SSH) and the Bonferroni approximation (BON) differ in how the joint budget $\epsilon$ is split among the reservoirs. BON splits the risk budget equally among reservoirs at each time step, whereas SSH is more flexible and able to dynamically allocate the risk budget when different units are strained. Fig.~\ref{fig:neighbors} shows how many hours of the water year each framework meets the joint reliability target. Note that an infeasible hour does not mean the optimization solver failed, but indicates that the controller could not operate $1-\epsilon$. SSH meets the target more often at every reliability level.

\subsubsection{Risk Attitude Analysis}

\begin{figure}[t]
    \centering
\includegraphics[width=0.9\columnwidth]{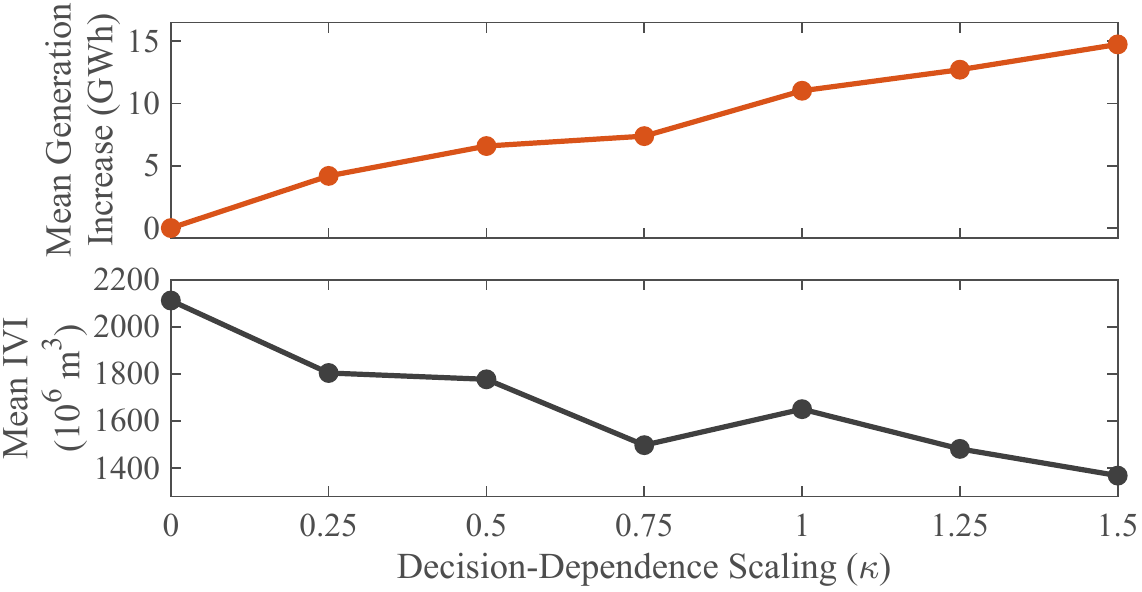}
    \caption{Average generation increase (top) and system-wide constraint violation (bottom) across a scaled DDU upstream-release coefficient, averaged over inflow scenarios with $\epsilon = 0.05$ and $\theta = 400$. Note that $\kappa=0$ recovers the DIU policy.}
    \label{fig:ddu-sensitivity}
\end{figure}

Fig.~\ref{fig:ddu-sensitivity} shows the sensitivity of system performance to the ARCH-X parameter $\gamma$, by applying a scaling factor $\kappa$ such that the model encodes $\kappa\gamma$, which controls the contribution of upstream release to downstream forecast variance. As $\kappa$ increases, the DDU model assigns greater uncertainty to upstream release behavior. This corresponds to more conservative reservoir operations and higher maintained forebay elevations. Consequently, average energy generation increases while system-wide constraint violations decreases over all scenarios. This behavior demonstrates how decision-dependent uncertainty promotes risk-aware water conservation without requiring explicit long-horizon optimization.

\subsubsection{Scalability}

Scalability is tested by tiling the four-unit cascade into synthetic systems with eight and twelve units. Each system is simulated for test year WY2024 under DDU with $\varepsilon=0.05$ and $\theta = 400$. Solve times scale by $n^{2.49}$ for SSH and  by $n^{1.19}$ for Bonferroni. Bonferroni is the faster algorithm because it solves a single linear program with a closed-form margin, while SSH re-solves once per cut iteration. At twelve units, SSH requires an average of $1.89$\,s to solve each rolling
horizon, well within the one-hour decision interval. Solve times differ from Table~\ref{tab:bench} because the scalability runs were executed serially while the benchmark runs shared the machine across seven parallel workers.

\begin{table}[t]
\centering
\caption{Computational cost against cascade size.}
\label{tab:scale}
\footnotesize
\setlength{\tabcolsep}{10pt}
\begin{tabular}{@{}rrrrrrr@{}}
\toprule
& & \multicolumn{3}{c}{SSH} & \multicolumn{2}{c}{BON} \\
\cmidrule(lr){3-5}\cmidrule(lr){6-7}
$n$ & Vars. & Iters. & Solve & Total & Solve & Total \\
& & & (ms) & (min) & (ms) & (min) \\
\midrule
$4$  & $960$  & $11.2$ & $121$  & $21.6$  & $24$ & $4.1$  \\
$8$  & $1920$ & $29.4$ & $635$  & $102.0$ & $59$ & $9.5$  \\
$12$ & $2880$ & $52.9$ & $1889$ & $290.1$ & $89$ & $14.4$ \\
\bottomrule
\end{tabular}
\end{table}

\section{Conclusion}\label{conclusion}

This paper develops a real-time coordination framework for cascaded hydropower systems under decision-dependent uncertainty. We explicitly model how upstream release decisions reshape downstream forecast variance while preserving tractable real-time dispatch. We formulate the problem as a rolling-horizon joint chance-constrained optimization program and develop a sequential supporting hyperplane algorithm to solve it. Numerical studies based on Lower Columbia River streamflow data demonstrate that incorporating decision-dependent uncertainty reduces the integrated violation index by up to 34.2\% and increases total generation by up to 12.8 GWh relative to DIU over a year of tested settings.

The proposed framework addresses the growing need for uncertainty-aware coordination of cascaded hydropower systems. Increasing climate variability and market participation require operators to continuously balance reliability, water conservation, and energy production under uncertain streamflow conditions. The proposed DDU framework enables more adaptive and risk-aware reservoir operation and can complement longer-term planning by updating release-dependent inflow uncertainty during real-time dispatch. Future work will extend the framework to larger reservoir networks, seasonal planning goals, and integration with electricity market participation under real-time bidding. These directions will be used to support more active hydropower participation in modern electricity markets while improving the resilience and flexibility of low-carbon power systems.

\bibliographystyle{IEEEtran}
\bibliography{IEEEabrv, main}

\appendix

\subsection{Derivation of the Bonferroni Reformulation}\label{app:bon}

The Bonferroni reformulation replaces the joint chance constraint~\eqref{g-volume} with $2n$ individual chance constraints, assigning equal risk $\varepsilon_i^{\mathrm{BON}}=\varepsilon/(2n)$ to the lower and upper volume constraints at each time step. By the Boole-Bonferroni inequality~\cite{Prekopa1988-he}, satisfying the
individual constraints guarantees the joint chance constraint, since

\begin{equation}
\mathbb P\Big[\bigcup_{i\in\mathcal I} \{v_{i,t}<\underline V_i\} \cup \{v_{i,t}>\overline V_i\} \Big] \le \sum_{i\in\mathcal I} 2\varepsilon_i^{\mathrm{BON}}
= \varepsilon
\end{equation}

Using the volume mass balance equation~\eqref{g-massbalance}, each volume bound can be written equivalently as a bound on the inflow, $\underline V'_{i,t}\le \hat q_{i,t}\le \overline V'_{i,t}$, such that
\begin{subequations}
\begin{align}
& \underline V'_{i,t} = \frac{1}{\kappa_{\mathrm{af}}}(\underline V_i-v_{i,t-1}) + u_{i,t}+s_{i,t} \\
& \overline V'_{i,t} = \frac{1}{\kappa_{\mathrm{af}}}(\overline V_i-v_{i,t-1}) + u_{i,t}+s_{i,t}
\end{align}
\end{subequations}

Since $\hat q_{i,t}\sim\mathcal N(\mu_{i,t},\hat\sigma_{i,t}^2)$, where
$\hat\sigma_{i,t}^2=(\boldsymbol{\Sigma}_t)_{ii}$ is the $i$th diagonal entry of $\Sigma_t$, the individual chance constraints have the deterministic equivalents~\cite{Prekopa1995-zq}

\begin{subequations}
\begin{align}
\mathbb P\big[\hat q_{i,t}\ge \underline V'_{i,t}\big]
&=
\Phi\left(
\frac{\mu_{i,t}-\underline V'_{i,t}}{\hat\sigma_i}
\right)
\ge 1-\varepsilon_i^{\mathrm{BON}} \\
\mathbb P\big[\hat q_{i,t}\le \overline V'_{i,t}\big]
&=
\Phi\left(
\frac{\overline V'_{i,t}-\mu_{i,t}}{\hat\sigma_i}
\right)
\ge 1-\varepsilon_i^{\mathrm{BON}} 
\end{align}
\end{subequations}

Letting $z=\Phi^{-1}(1-\varepsilon_i^{\mathrm{BON}})$ gives the linear inflow bounds
\begin{equation}
\underline V'_{i,t}+z\hat\sigma_i \le
\mu_{i,t} \le \overline V'_{i,t}-z\hat\sigma_i
\end{equation}

Substituting back the shifted bounds gives the effective volume constraints for the Bonferroni approximation.

\subsection{Proof of Theorem~\ref{the1}}\label{thm-convergence}

The proof is organized into the following three steps.

\textit{Step 1: Convexity and monotonic refinement.}
Since $\hat{\boldsymbol{q}}_t \sim \mathcal{N}(\boldsymbol{\mu}_t, \boldsymbol{\Sigma}_t^{\mathrm{DDU}})$ is multivariate Gaussian, the chance-constrained feasible set
\begin{equation}
    \Omega_t = \{\boldsymbol{x}_t : F(\underline{\boldsymbol{V}}'_t(\boldsymbol{x}_t), \overline{\boldsymbol{V}}'_t(\boldsymbol{x}_t); \boldsymbol{\mu}_t, \boldsymbol{\Sigma}_t^{\mathrm{DDU}}) \ge 1 - \varepsilon\}
\end{equation}
is convex by the log-concavity of the multivariate Gaussian CDF~\cite{Prekopa1978-bo}. The polyhedral approximations $\mathcal{F}_t^{(k)}$ to $\Omega_t$, generated by the SSH algorithm via cutting planes, satisfy the monotonic nesting:
\begin{equation}\label{nesting-sets}
    \Omega_t \subseteq \mathcal{F}_t^{(K)} \subseteq \cdots \subseteq \mathcal{F}_t^{(1)} \subseteq \mathcal{F}_t^{(0)}
\end{equation}
where $\mathcal{F}_t^{(0)} = \{\boldsymbol{x}_t : \boldsymbol{A}_t \boldsymbol{x}_t \le \boldsymbol{b}_t\}$ is the deterministic linear feasible region of~\eqref{GES} obtained by ignoring~\eqref{g-volume}. The convexity of $\Omega_t$ makes supporting hyperplanes well-defined at any boundary point, which we exploit in Step 2 to refine $\mathcal{F}_t^{(k)}$.

\textit{Step 2: Hyperplane separation.}
If $\boldsymbol{x}_t^{(k)} \notin \Omega_t$, i.e., $F(\boldsymbol{x}_t^{(k)}) < 1-\varepsilon$, the line search between $\boldsymbol{x}_t^{(k)}$ and the Slater point $\boldsymbol{x}_t^s$ yields, by continuity of $F$ and the intermediate value theorem, a point $\boldsymbol{x}_t^{(k,*)} = (1-\lambda^*)\boldsymbol{x}_t^{(k)} + \lambda^* \boldsymbol{x}_t^s$ with $\lambda^* \in (0, 1]$ such that $F(\boldsymbol{x}_t^{(k,*)}) = 1-\varepsilon$. Because $F$ is log-concave, $g_t(\boldsymbol{x}_t)=\log(1-\varepsilon)-\log F(\boldsymbol{x}_t)$ is convex and
$\Omega_t=\{\boldsymbol{x}_t:g_t(\boldsymbol{x}_t)\leq0\}$.
The supporting hyperplane of $g_t$ at
$\boldsymbol{x}_t^{(k,*)}$ is equivalently
\begin{equation}
    -\nabla F(\boldsymbol{x}_t^{(k,*)})^\top \boldsymbol{x}_t \le -\nabla F(\boldsymbol{x}_t^{(k,*)})^\top \boldsymbol{x}_t^{(k,*)}
\end{equation}
which separates $\boldsymbol{x}_t^{(k)}$ from $\Omega_t$. Adding this hyperplane to $\mathcal{F}_t^{(k)}$ excludes $\boldsymbol{x}_t^{(k)}$ from $\mathcal{F}_t^{(k+1)}$, ensuring strict refinement. Since each iteration strictly shrinks the polyhedral approximation, the sequence $\{\boldsymbol{x}_t^{(k)}\}$ cannot revisit previously infeasible iterates, setting up the convergence argument in Step 3.

\textit{Step 3: Finite termination.}
The strict refinement in Step 2, combined with the compactness of $\mathcal{F}_t^{(0)}$ (bounded by the variable bounds in~\eqref{GES}) and the continuity of $F$, ensures that consecutive iterates $\{\boldsymbol{x}_t^{(k)}\}$ approach each other monotonically. For any prescribed tolerance $\epsilon > 0$, there exists a finite $K$ such that $\|\boldsymbol{x}_t^{(K+1)} - \boldsymbol{x}_t^{(K)}\|_\infty < \epsilon$. By the nesting relation~\eqref{nesting-sets}, $\mathcal{F}_t^{(K)}$ contains $\Omega_t$, so $\boldsymbol{c}^\top \boldsymbol{x}_t^{(K)}$ provides a lower bound on the optimal value of~\eqref{GES}, and $\boldsymbol{x}_t^* = \boldsymbol{x}_t^{(K)}$ is $\epsilon$-optimal in the solution norm. This completes the proof. $\blacksquare$

\begin{figure}[t]
\centering
\subfloat[\label{fig:ssh-convergence}]{%
    \includegraphics[width=0.85\columnwidth]{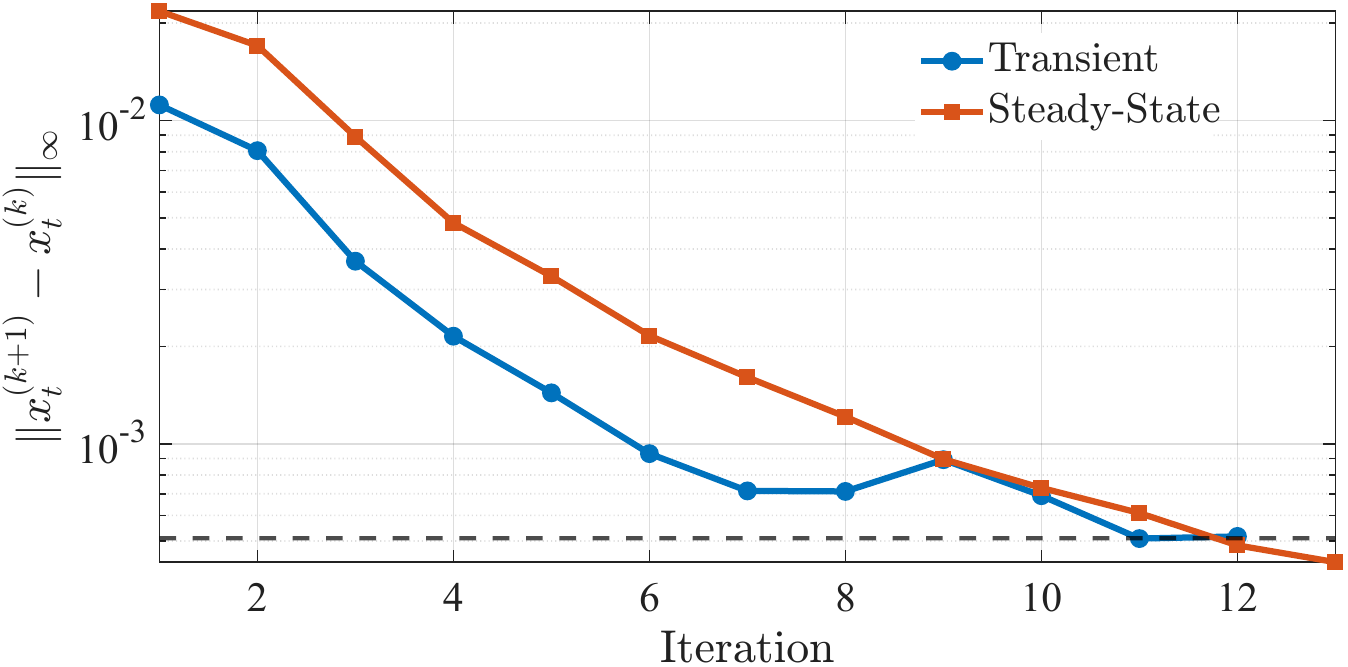}
} \\
\vspace{-1em}
\subfloat[\label{fig:risk}]{%
    \includegraphics[width=0.85\columnwidth]{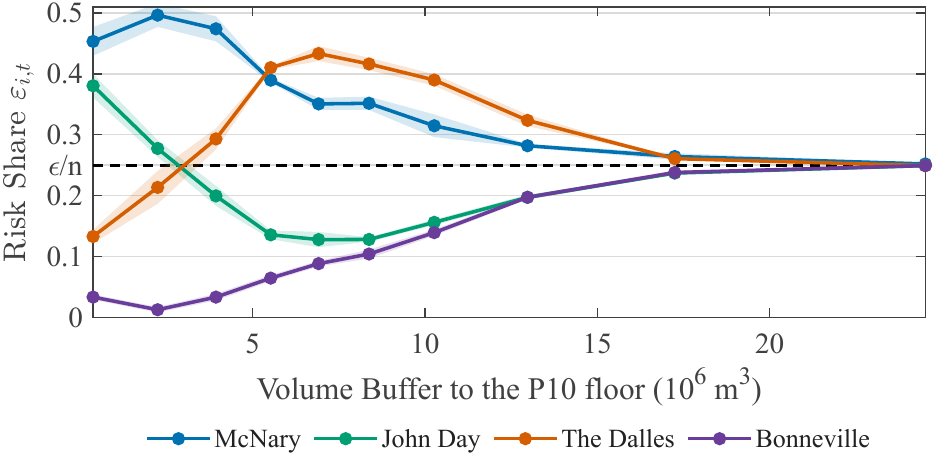}
}
\caption{ 
(a) Convergence trajectories of SSH under transient and steady-state operating conditions (b) Mean SSH risk allocation $\varepsilon_{i,t}$ for the four-unit cascade, where $\alpha_i$ denotes the per-unit share of the joint budget $\varepsilon$. Risk allocation traces from transient conditions to steady state conditions and converge at $\epsilon/n$.}
\label{fig:risk-converge}\vspace{-1em}
\end{figure}

Fig.~\ref{fig:ssh-convergence} shows SSH convergence under transient
conditions with varying stress across units and steady-state conditions
with uniform risk. Both cases converge in finitely many iterations as 
consistent with Theorem~\ref{the1}.

\subsection{Proof of Proposition~\ref{prop1}}\label{thm-risk}

The proof shows that symmetric steady-state conditions imply
equal gradient magnitudes across all units, yielding uniform risk
attribution $\alpha_{i,t}=1/n$.

\textit{Step 1: Define attribution via CDF gradient.}
Recall
$F(\underline{\boldsymbol{V}}'_t,\overline{\boldsymbol{V}}'_t;
\boldsymbol{\mu}_t,\boldsymbol{\Sigma}_t)$
from~\eqref{joint-dist} is the multivariate Gaussian probability that
all units simultaneously satisfy the volume bounds. By the chain
rule applied to the shifted bounds
$\underline{V}'_{i,t}
=\kappa_{af}^{-1}(\underline{V}_i-v_{i,t-1})+u_{i,t}+s_{i,t}$
and
$\overline{V}'_{i,t}
=\kappa_{af}^{-1}(\overline{V}_i-v_{i,t-1})+u_{i,t}+s_{i,t}$,
the sensitivity of $F_t$ to the state $v_{i,t-1}$ satisfies
\begin{equation}\label{vol-risk}
    \frac{\partial F_t}{\partial v_{i,t-1}}
    =
    -\frac{1}{\kappa_{af}}
    \left(
    \frac{\partial F_t}{\partial \underline{V}'_{i,t}}
    +
    \frac{\partial F_t}{\partial \overline{V}'_{i,t}}
    \right).
\end{equation}
The SSH risk attribution signal is defined as the normalized magnitude:
\begin{equation}\label{risk-signal}
    \alpha_{i,t}
    =
    \frac{\left|\partial F_t/\partial v_{i,t-1}\right|}
    {\sum_{j=1}^{n}
    \left|\partial F_t/\partial v_{j,t-1}\right|}.
\end{equation}

\textit{Step 2: Establish gradient symmetry.}
The joint distribution
$\hat{\boldsymbol{q}}_t\sim
\mathcal{N}(\boldsymbol{\mu}_t,\boldsymbol{\Sigma}_t)$
and bounds
$\underline{\boldsymbol{V}},\overline{\boldsymbol{V}}$
are invariant under symmetric steady-state conditions. Since $F_t$ depends only on these quantities, $F_t$ is
also permutation-invariant, hence
\begin{equation*}
    \left|\frac{\partial F_t}{\partial v_{i,t-1}}\right|
    =
    \left|\frac{\partial F_t}{\partial v_{j,t-1}}\right|,
    \quad \forall i,j\in\mathcal I.
\end{equation*}
For $\boldsymbol{\Sigma}_t\succ0$, $F_t$ is smooth; we assume
$\nabla F_t\neq0$ at the boundary points evaluated by SSH.

\textit{Step 3: Conclude uniform attribution.}
Substituting the equal gradients into~\eqref{risk-signal} yields
\begin{equation*}
    \alpha_{i,t}
    =
    \frac{|\partial F_t/\partial v_{i,t-1}|}
    {n|\partial F_t/\partial v_{i,t-1}|}
    =\frac{1}{n},
    \quad \forall i\in\mathcal I.
\end{equation*}

\noindent Therefore,
$\varepsilon_{i,t}^{\mathrm{SSH}}
=\varepsilon\alpha_{i,t}=\varepsilon/n$.
This establishes uniform risk attribution under symmetric steady-state
conditions. $\blacksquare$

Fig~\ref{fig:risk} shows the corresponding SSH risk allocations from real-time simulations. Risk is dynamically shifted toward the most constrained unit and returns to a uniform distribution as conditions stabilize, consistent with Proposition~\ref{prop1}.

\end{document}